# Excitation of spin waves by spin polarized current.

K. Rivkin[1] and J. B. Ketterson[1,2,3]

1. Department of Physics and Astronomy, Northwestern University

Evanston IL, 60201

2. Department of Electrical and Computer Engineering, Northwestern University

Evanston IL, 60201

3. Materials Research Center, Northwestern University

Evanston IL, 60201

**Abstract**

Numerical and analytical analysis is used to explain recently observed experimental phenomenon – excitation of spin waves in spin valves due to the applied spin polarized current. Excited spin waves are being identified and stability analysis is being used to identify different regimes of spin waves' excitation depending on the value of the applied current.

In a recently introduced new branch of electronics, called spintronics[1] the operation of individual devices is due to their interaction with a spin polarized current. In particular, analogously to the conventional magnetic memory systems, operated by external magnetic fields, one can use spin polarized current in order to produce magnetization reversal[2]. While the importance of spin waves in operating spin valves have been long known[3], the excitation of spin waves by the spin polarized current only recently have been observed by Kiselev et al[4]. In their experiment even through the current strength was below the value required for the reversal, a complex oscillation of magnetic moments was induced in the sample. In a later article by Lee et al[5] it was explained that these oscillations correspond to the excitation of inhomogeneous spin waves by the applied spin polarized current. However, individual excited modes have not been identified and the detailed analysis of their excitation have not been performed.

In the present work we will attempt to contribute to the further understanding of this phenomenon.

We recall the Landau-Lifshitz[6] equation in the presence of a dissipative Gilbert and Spin Transfer Torque[7,8,9,10,11] terms:

$$\frac{d\mathbf{m}}{dt} = -\gamma \mathbf{m} \times \mathbf{h} ; \qquad (1)$$

$$\mathbf{h} = \mathbf{h}^{true} + \frac{\mathbf{m}}{M_s} \times \left( \beta \mathbf{h}^{true} - I \mathbf{h}^J \right) \qquad (2)$$

here $\mathbf{m}$ is the magnetic moment, $\gamma$ is the gyromagnetic ratio, and $\beta$ is a parameter governing the dissipation; the interaction with spin polarized current is being described by the term $\frac{\mathbf{m}}{M_s} \times \left( \beta \mathbf{h}^{true} - I \mathbf{h}^J \right)$, where $\mathbf{h}^J$ is the polarization of the current and I is an empirical factor measuring the strength of the coupling (in units of magnetic field where 1000 Oe corresponds to $10^8$ A/cm$^2$).

Typical spin valve consists of two magnetic layers, separated by a nonmagnetic layer. One of the layers is assumed to have a constant magnetization, and another one, so called free layer, has a magnetic moment that can change with time. Initially both layers are magnetized along the same direction; spin polarized current when it enters the free layer is also polarized along the same direction.

Here we will use parameters similar to those used by Kiselev et al[4,12] in which the free layer is 2nm thick and another layer is 10nm thick. The layers are deposited from Cobalt and approximate an ellipse with principal diameters 130 by 70nm. We assume there is no exchange interaction acting between the layers. We take exchange stiffness $A = 1.3 \cdot 10^{-6}$ erg/cm and saturation magnetization $M_S = 795$ emu/cm$^3$. This value is very different from the one which was used by Kiselev[4] in order to fit the frequency of the uniform mode. It is our opinion that if the numerical eigenvalue algorithm, as described below, or Kittel's formula with numerically calculated demagnetization tensor components is being used, the uniform mode frequency is going to be more consistent with experimental results.

The coordinate system is chosen such that the easy axis are parallel to the z-axis with the magnetic layers lying in the y-z plane; the current direction is along x direction. The current is polarized along z axis. In our case a current of 1 mA should correspond to $a_J \cong 144$ Oe, and the damping coefficient β is taken as[4] 0.014. Two cases will be considered – one in which the free layer is subjected to the magnetic field of the thick layer and external magnetic field $H_o$=2600 Oe. In the second case it is assumed that the external magnetic field cancels out the magnetic field of the thick layer.

We start with linearizing Landau-Lifshitz equation with respect to time dependent oscillating part of magnetization $\mathbf{m}_i^{(1)}(t) = \mathbf{m}_i^{(1)} e^{-i\omega t}$ and corresponding oscillating magnetic fields $\mathbf{h}_i^{(1)}(t)$. We assume that initially the system is in equilibrium with magnetic moments $\mathbf{m}_i^{(0)}$ aligned parallel to local magnetic fields $\mathbf{h}_i^{(0)}$. In the absence of current, we have the following eigenvalue equation[13]:

$$-\frac{d\mathbf{m}_i^{(1)}}{dt} = \gamma \left[ \mathbf{m}_i^{(0)} \times \mathbf{h}_i^{(1)} + \mathbf{m}_i^{(1)} \times \mathbf{h}_i^{(0)} \right] + \frac{\gamma \beta}{M_s} \mathbf{m}_i^{(0)} \times \left[ \mathbf{m}_i^{(0)} \times \mathbf{h}_i^{(1)} + \mathbf{m}_i^{(1)} \times \mathbf{h}_i^{(0)} \right]$$
$$i\omega_k \mathbf{V}_i^{(k)} = \gamma \mathbf{B}_{ij} \mathbf{V}_j^{(k)}$$
(3)

where $\mathbf{V}_i^{(k)}$ and $\omega_k$ are eigenvectors and eigenvalues. Solving Eq.(3) provides us with an entire spectrum of spin waves present in the sample.

What happens if a spin polarized current is being applied to the sample? In this case Eq.(3) transforms into an inhomogeneous equation with a source term depending on the applied current:

$$\frac{d\mathbf{m}^{(1)}}{dt} + \gamma \left[ \mathbf{m}^{(1)} \times \mathbf{h}^{(0)} + \mathbf{m}^{(0)} \times \mathbf{h}^{(1)} \right] + \frac{\gamma \beta}{M_s} \mathbf{m}_i^{(0)} \times \left[ \mathbf{m}_i^{(0)} \times \mathbf{h}_i^{(1)} + \mathbf{m}_i^{(1)} \times \mathbf{h}_i^{(0)} \right] \approx -\gamma \mathbf{g}(t)$$
$$\mathbf{g}(t) = -\frac{I}{M_s} \mathbf{m}^{(0)} \times (\mathbf{m}^{(0)} \times \mathbf{h}^J) - \frac{I}{M_s} \mathbf{m}^{(0)} \times (\mathbf{m}^{(1)} \times \mathbf{h}^J) - \frac{I}{M_s} \mathbf{m}^{(1)} \times (\mathbf{m}^{(1)} \times \mathbf{h}^J)$$
(4)

here we neglected the terms like $\frac{I}{M_s} \mathbf{m}^{(1)} \times (\mathbf{m}^{(0)} \times \mathbf{h}^J)$ - it can be shown that they do not contribute to the dynamics of spin waves; $\mathbf{g}(t)$ is a source term responsible for the excitation of spin waves..

The solution of this equation will consist of two parts – solution of a homogeneous Eq.(3) and solution of inhomogeneous Eq.(4). The sum of these solutions should be chosen in such a way as to satisfy the initial conditions. Since we are mostly interested in steady-state solutions, we will not analyze in detail the initial conditions and solutions of Eq.(3) – in the presence of damping ($\beta > 0$) all of such solutions are evanescent in time and do not contribute to the steady state behavior.

While mathematically rigid solution of Eq.(4) have been provided in[14], here we will use a less rigid but nevertheless correct method of analysis, first introduced in[15]. Suppose that the solution of Eq.(4) can be expanded into solutions of a homogeneous Eq.(3):

$$\mathbf{m}^{(1)} = \sum_k a_k(t)\mathbf{V}^{(k)} \tag{5}$$

In this case Eq.(4) becomes:

$$\frac{da_k(t)}{dt} + ia_k(t)\omega_k = -\gamma \iiint \mathbf{g}\mathbf{V}^{*(k)}dxdydz \tag{6}$$

in a discrete version integration is obviously going to be replaced by summation. Starting with the first part of the source term in Eq.(4):

$$\frac{da_k(t)}{dt} + ia_k(t)\omega_k = \gamma \frac{I}{M_s} \sum \left(\mathbf{m}^{(0)} \times (\mathbf{m}^{(0)} \times \mathbf{h}^J)\right) \cdot \mathbf{V}^{*(k)}$$

$$a_k(t) = \gamma I \frac{\sum \left(\mathbf{m}^{(0)} \times (\mathbf{m}^{(0)} \times \mathbf{h}^J)\right) \cdot \mathbf{V}^{*(k)}}{i\omega_k M_s} \tag{7}$$

As one can see in this case the solution is time independent and simply represents a constant difference from the initial equilibrium. $a_k$ in this case is nonzero only for the modes that are concentrated in the areas where equilibrium magnetization is not parallel to the current's polarization and therefore is not parallel to the external field. This means that the modes excited by this term are going to be edge modes; evaluating the term on the right side of Eq.(7) tells us that, for example, if $H_0$=2600 Oe, the mode with frequency f=21.79 GHz (as we mentioned before, solving Eq.(3) gives us all the spin waves and their frequencies) is going to be nearly the only mode excited by such term. Its spatial distribution is given in Fig. 10.f. The excitation of this mode corresponds to the static difference between the new equilibrium, formed in the presence of current, and the old equilibrium.

Second term in Eq.(4) provides:

$$\frac{da_k(t)}{dt} + ia_k(t)\omega_k = \gamma \frac{I}{M_s} \sum \left( \mathbf{m}^{(0)} \times (\sum_{k'} a_{k'}(t) \mathbf{V}^{(k')} \times \mathbf{h}^J) \right) \cdot \mathbf{V}^{*(k)} \quad (8)$$

$$\frac{da_k(t)}{dt} + ia_k(t)\omega_k = \gamma \sum_{k'} c_{kk'} a_{k'}(t)$$

$$c_{kk'} = \frac{I}{M_s} \sum_{x,y,z} \mathbf{V}^{(k')} \mathbf{V}^{*(k)} \left( \mathbf{m}^{(0)} \cdot \mathbf{h}^J \right) \quad (9)$$

If all of the magnetic moments are parallel to each other $c_{kk'} = I\frac{m_z^{(0)}}{M_s}\delta_{kk'}$; since current's polarization is aligned with the easy axis and external field, this is almost the case - if k prime is not equal to k, $c_{kk'}$ initially can maximum be a few percent of the value $I\frac{m_z^{(0)}}{M_s}$. For the time we can assume that $c_{kk'} = I\frac{m_z^{(0)}}{M_s}\delta_{kk'}$ and:

$$\frac{da_k(t)}{dt} + ia_k(t)\omega_k = \gamma I \frac{m_z^{(0)}}{M_s} a_k(t) \quad (10)$$

$$a_k(t) = a_k(0)e^{i\alpha t} = a_k(0)e^{-i\omega_k t}e^{\gamma I \frac{m_z^{(0)}}{M_s} t} \quad (11)$$

Since $\omega_k = \omega_k' - i\omega_k''$:

$$a_k(t) = a_k(0)e^{-i\omega_k' t}e^{\left(\gamma I \frac{m_z^{(0)}}{M_s} - \omega_k''\right)t}$$

$$\gamma I_c \frac{m_z^{(0)}}{M_s} = \omega_k''$$
(12)

As one can see there are two options: if $\gamma I \frac{m_z^{(0)}}{M_s} < \omega_k''$ then the oscillations are dying out exponentially in time. If $\gamma I \frac{m_z^{(0)}}{M_s} > \omega_k''$ then the excitation of $k^{th}$ mode is going to exponentially increase with time.

Usually the uniform mode has the lowest frequency, both real and imaginary parts. Because of it, uniform mode most likely is going to be the most excited mode; with higher currents higher frequency modes are going to be excited, however the speed of their excitation is going to be smaller than that of a uniform mode.

It is obvious that exponential growth can not continue forever. One of the possibilities is that the term $\frac{I}{M_s}\mathbf{m}^{(1)} \times (\mathbf{m}^{(1)} \times \mathbf{h}^J)$ will create a highly non-linear coupling between the excited modes. In order to explain other two possibilities we need to emphasize that for sufficiently strong excitation, Eq.(5) does not account for the fact that the eigenvectors are time-dependent. While non-linear mode mixing depends on the excitation of higher-frequency modes, for now we consider only the processes that result from the excitation of the uniform mode. Because of this, we can use a macrospin approximation, with the shape of the sample responsible for the values of the demagnetization tensor $A_{\alpha\beta}$:

$$H_\alpha = \sum_\beta A_{\alpha\beta} M_\beta \tag{13}$$

For the above mentioned ellipse (130x70nm), demagnetization tensor has all zeroe off-diagonal elements, and the diagonal elements can be estimated numerically to be $A_{zz}=-0.27$, $A_{yy}=-0.61$, $A_{xx}=-11.69$. From studying the dynamics of the sample we also know that x component of the magnetization always remain small and therefore $m_x^{(0)}$ can be neglected in our approximate analysis.

Modifying Kittel's formula accounting for the presence of damping one has:

$$\omega = \gamma\sqrt{\left(H_0 + m_z^{(0)}B\right)\left(H_0 + m_z^{(0)}A\right)} - i\gamma\frac{\beta}{M_s} m_z^{(0)} \left(\frac{\left(H_0 + m_z^{(0)}B\right) + \left(H_0 + m_z^{(0)}A\right)}{2}\right) \tag{14}$$

where in our case

$$\begin{aligned}(A_z - A_y) &= A = -0.27 + 0.61 = 0.34 \\ (A_z - A_x) &= B = -0.27 + 11.69 = 11.42 \\ (A_y - A_x) &= C = -0.61 + 11.69 = 11.08\end{aligned} \tag{15}$$

in the absence of external field, Eq. (14) can be extended to the case of arbitrary direction of magnetization in z-y plane:

$$\omega = \gamma\sqrt{m_z^{(0)^2} AB - m_y^{(0)^2} AC} - i\gamma\frac{\beta}{2M_s}\left(m_z^{(0)^2}(A+B) + m_y^{(0)^2}(C-A)\right) \tag{16}$$

Inserting this expression into Eq.(12) we can see that a steady state solution is possible (for the first time the existence of such solution was predicted by Li and Zhang[10]): while at t=0 the current is larger than the imaginary part of the uniform mode's frequency, for certain value of $m_z^{(0)}$ the following expression is satisfied:

$$\gamma I_c \frac{m_z^{(0)}}{M_s} = \omega_k''$$

$$\frac{\beta}{2}\left(m_z^{(0)^2}(A+B) + m_y^{(0)^2}(C-A)\right) = I m_z^{(0)} \tag{17}$$

and the uniform mode simply oscillates with its own resonant frequency.

However, this is not the only possible behavior. Since the eigenvector depends on the direction of magnetization as:

$$V_z = \frac{m_y^{(0)} C}{\sqrt{m_z^{(0)^2} AB - m_y^{(0)^2} AC + m_z^{(0)^2} B^2 + m_y^{(0)^2} C^2}}$$

$$V_x = \frac{i\sqrt{m_z^{(0)^2} AB - m_y^{(0)^2} AC}}{\sqrt{m_z^{(0)^2} AB - m_y^{(0)^2} AC + m_z^{(0)^2} B^2 + m_y^{(0)^2} C^2}} \tag{18}$$

$$V_y = = \frac{m_z^{(0)} B}{\sqrt{m_z^{(0)^2} AB - m_y^{(0)^2} AC + m_z^{(0)^2} B^2 + m_y^{(0)^2} C^2}}$$

it is clear that initially ($m_y^{(0)} = 0$) the oscillations are confined to x-y plane; excitation of the uniform mode leads to slow decrease of $m_z^{(0)}$, however when $m_z^{(0)}$ is small the oscillations are confined mostly to z-y plane. Because of this, even if the system reaches the value of $m_z^{(0)}$ which satisfies Eq.(17), the oscillations in z-y plane do not stop. If the current is not strong enough, it is possible that the magnetization will move towards smaller values of $m_z^{(0)}$, oscillations will increase, than the magnetization will move towards larger of values of $m_z^{(0)}$ and the system will enter the region where the damping exceeds the excitation due to the current, and so on, back and forth, produce complex oscillatory pattern. In this case non-linear interaction between excited modes should also be taken into account. However, if the current is strong enough, and therefore the value of $m_z^{(0)}$ for which the damping exceeds the excitation is small enough, the value of $m_z^{(0)}$ becomes negative during one of such oscillations[8]. As soon as it happens, the term in Eq.(12) due to the applied current changes sign and acts as additional damping. The more negative $m_z^{(0)}$ is, the higher the damping is, and, as a result of it, the system switches to the new equilibrium configuration, with the magnetization aligned in the opposite direction with respect to the initial configuration.

Since we have made numerous approximations in our analytical treatment of the problem, we need to verify our results at least by performing a numerical experiment. We will use Runge-Kutta simulation: after obtaining the equilibrium configuration in the absence of the current we will randomly slightly perturb the magnetic moments in the system, simulating a temperature excitation of the system, while in the same time applying the spin polarized current. We will than proceed with taking Fourier Transform of time dependency of $M_z$, which corresponds to experimental measurements of the sample's resistance[4]. However, Fourier analysis of one of the magnetization components is not going to tell us which modes are excited – growing and decaying exponents are going to broaden the spectrum corresponding to the excited modes and some of the effects we did not address in detail in this work can also drastically affect the Fourier spectrum. In order to have the detailed analysis of the modes' excitation we should spatially decompose the oscillations in the system into the spectrum of spin waves:

$$\begin{aligned} \mathbf{m}^{(1)}(t) &= \mathbf{m}(t) - \mathbf{m}^{(0)} \\ a_k(t) &= \iiint \mathbf{m}^{(1)}(t) \cdot \mathbf{V}^{(k)} dxdydz \end{aligned} \qquad (19)$$

We start with a "damped" regime, in which most of the oscillations should die out according to Eq.(12). At first we see a small excitation of nearly all modes, which can be explained by the fact that the momentarily applied current possesses a broad Fourier spectrum, therefore it initially able to excite a broad spectrum of modes. Following this, all of the modes die out exponentially, with an exception of the modes identified from Eq.(7). The numerically calculated threshold current value typically a little bit, 0.07 mA or so, higher than the one given by Eqs. (12) and (14). For the currents above this value (0.49 mA with no external field, 0.71 mA for $H_0$=2600 Oe) we indeed see a steady-state behavior, as shown on Fig.1. As it was said before, the oscillations of z component of magnetization are determined mostly by z component of the eigenvector (Fig. 2). If the current value is increased, a complex oscillation pattern forms, both in oscillations of the uniform mode (Fig.3) and z component of the magnetization (Fig. 4). If the current is increased even more, switching occurs (Figs. 5 and 6). As one can see the general behavior is consistent with the analytical analysis – the uniform mode grows until a semi-steady state forms for a very small value of $M_z$, then due to non-zero value of $V_z$ the system acquires a small but negative value of $M_z$. As soon as it happens, the modes are quickly become suppressed by the combined force of damping and applied current.

It is interesting to compare our numerical experiments with the real experiments, like the ones described in[4]. Assuming that the external field applied to the spin valve is $H_0$=2600 Oe, we repeat the calculations. The results are basically the same as before, Fig. 7 shows the excitation of the uniform mode for the oscillatory regime, Fig. 8 shows the Fourier spectra of $M_z(t)$. It has a strong peak in the vicinity of 23 GHz, however, which modes are excited can not be known from the Fourier spectrum alone: most of the modes excited in the system initially have frequencies below 20 GHz (Fig.10); moreover, in general, their frequencies decrease with the decrease of $M_z$. Fig. 9 shows the relative maximum excitation of modes for different currents, corresponding respectively to damped, steady state and oscillatory regimes. As one can see in a damped regime excitation of modes is rather non-selective, and determined by the coupling between initial random excitations and the momentarily applied current. In the steady state regime only a very few low frequency modes satisfy Eq. (12) and as a result, only few modes are excited. In an oscillatory regime, the current is strong enough to excite many modes, however, low frequency modes are excited stronger than higher frequency ones – the exponent in Eq.(12) is typically larger for low than high frequency modes. The spatial distribution of the first six modes is given in Fig. 10. Only out-of-plane component of magnetization is being presented, with colors corresponding to different strengths of excitation.

*Conclusions*

In our work we succeeded in establishing the following: there are four possible regimes of excitation of spin waves in the presence of the applied spin polarized current, depending on the current's value:

      a. Damped regime, where simply a new static equilibrium is attained.

      b. Steady-state regime, characterized by nearly constant excitation of a uniform mode and relatively small excitation of other modes.

      c. Oscillatory regime, where the broad spectrum of modes is being excited.

d. Switching regime, where the semi-steady state excitation of modes is superseded by very fast damping to the new static equilibrium, with the magnetization anti-parallel to its initial value.

**Acknowledgments.**

This work was supported by the National Science Foundation under grants ESC-02-24210 and DMR 0244711. The software used in making this article is freely available at www.rkmag.com. We would like to express our gratitude to Professor D. Ralph of Cornell University for his invaluable comments and advice.

## References

[1] J. Slonczewski Journal of Magnetism and Magnetic Materials **159**, L1 (1996).

[2] J. Glorrier et al, Apllied Physics Letters **78**, 3663 (2001).

[3] R. Jansen et al, Physical Review Letters **85**, 3277 (2000).

[4] S. I. Kiselev, J. C. Sankey, I. N. Krivorotov, N. C. Emley, R.J. Schoelkopf, R.A. Buhrman and D. C. Ralph Nature **425**, 380 (2003).

[5] Kyung-Jin Lee et al, Nature Materials **3**, 877 (2004).

[6] L. D. Landau and E. M. Lifshitz, Phys. Z. Soviet Union **8**, 153 (1935).

[7] A. Brataas, Y.V. Nazarov and G. E. Bauer, Physical Review Letters **84**, 2481 (2000).

[8] S. Zhang, P. M. Levy and A. Fert, Physical Review Letters **88**, 236601 (2002).

[9] J. E. Wegrove et al Europhysics Letters **45**, 626 (1999).

[10] Z. Li and S. Zhang Physical Review B **68**, 024404 (2003).

[11] J. Z. Sun Physical Review B **62**, 570 (2000).

[12] S. I. Kiselev, J. C. Sankey, I. N. Krivorotov, N. C. Emley, A. G. F. Garcia, R. A. Buhrman and D. C. Ralph "Spin Transfer Excitations for permalloy nanopillars for large applied currents", http://www.arxiv.org/list/cond-mat/0504402.pdf

[13] K. Rivkin, A. Heifetz, P. R. Sievert and J. B. Ketterson, "Resonant modes of dipole-coupled lattices", Physical Review **B 70**, 184410 (2004).

[14] K. Rivkin and J.B.Ketterson, submitted to Physical Review B.

[15] K. Rivkin, L.E. DeLong and J.B. Ketterson, "Microscopic study of magnetostatic spin waves", Journal of Applied Physics **97**, 10E309 (2005).

**Figures.**

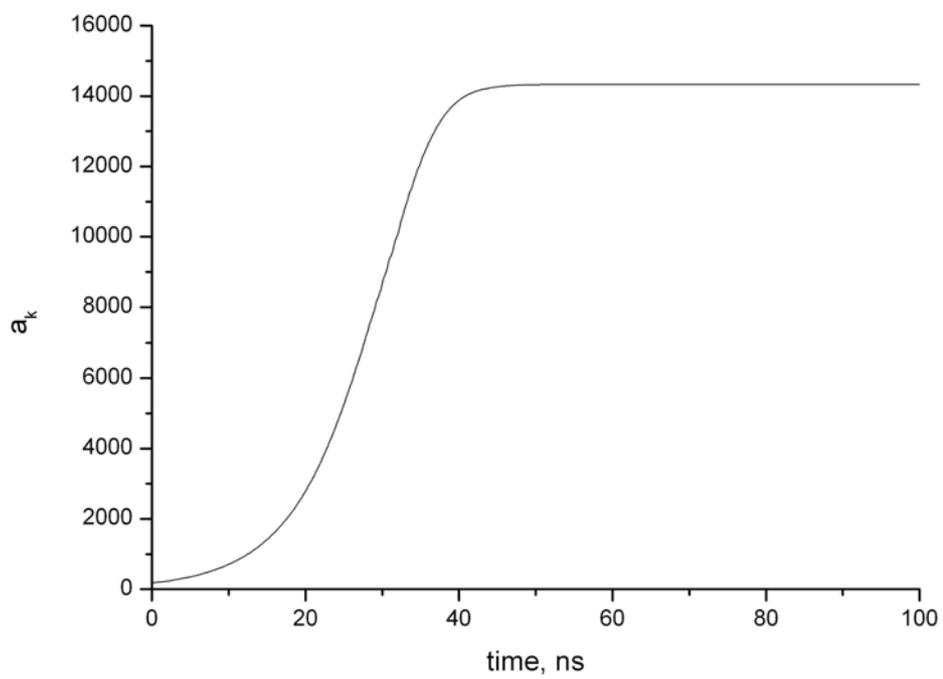

**Figure 1. Amplitude of the uniform mode as a function of time, no external field, current amplitude 0.5 mA.**

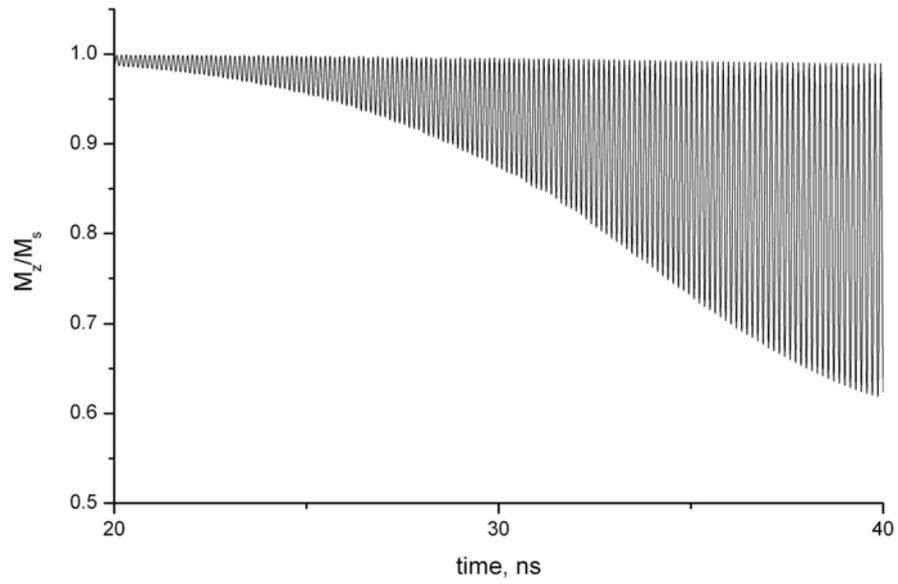

**Figure 2. Z component of the sample's magnetization as a function of time, no external field, current amplitude 0.5 mA.**

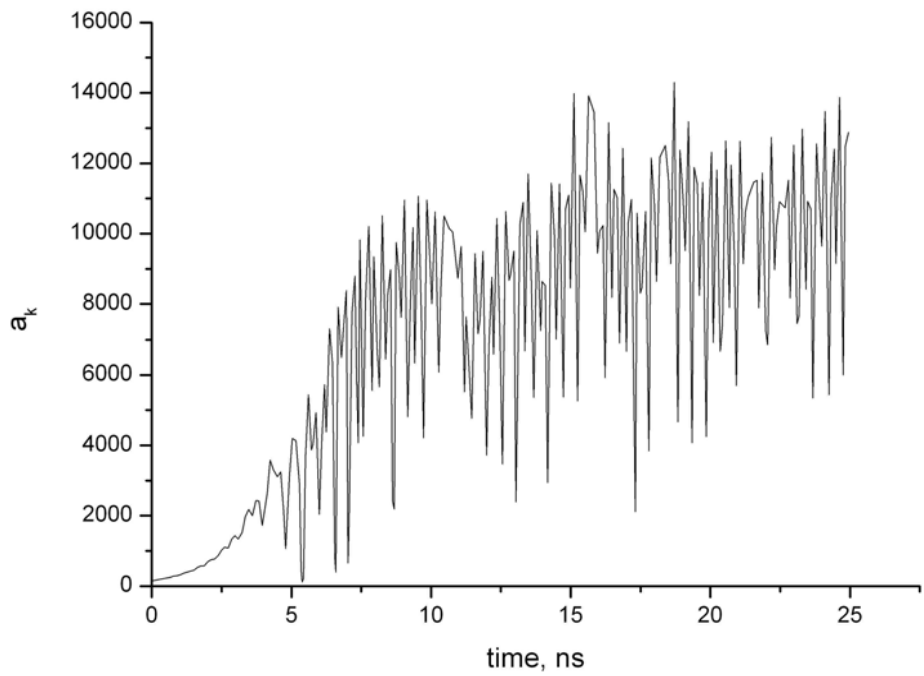

**Figure 3. Amplitude of the uniform mode as a function of time, no external field, current amplitude 0.74 mA.**

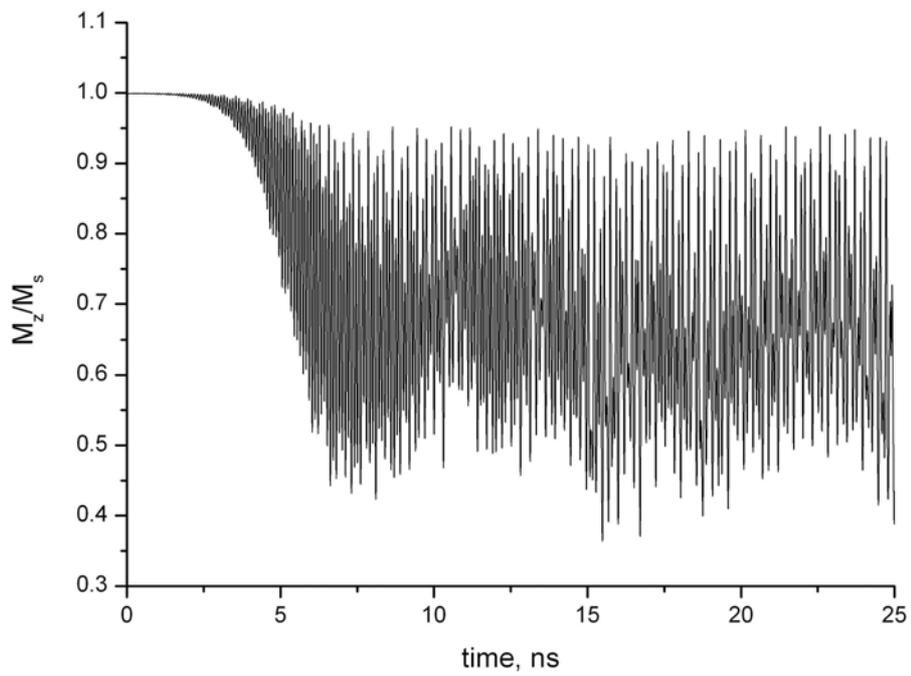

**Figure 4. Z component of the sample's magnetization as a function of time, no external field, current amplitude 0.74 mA.**

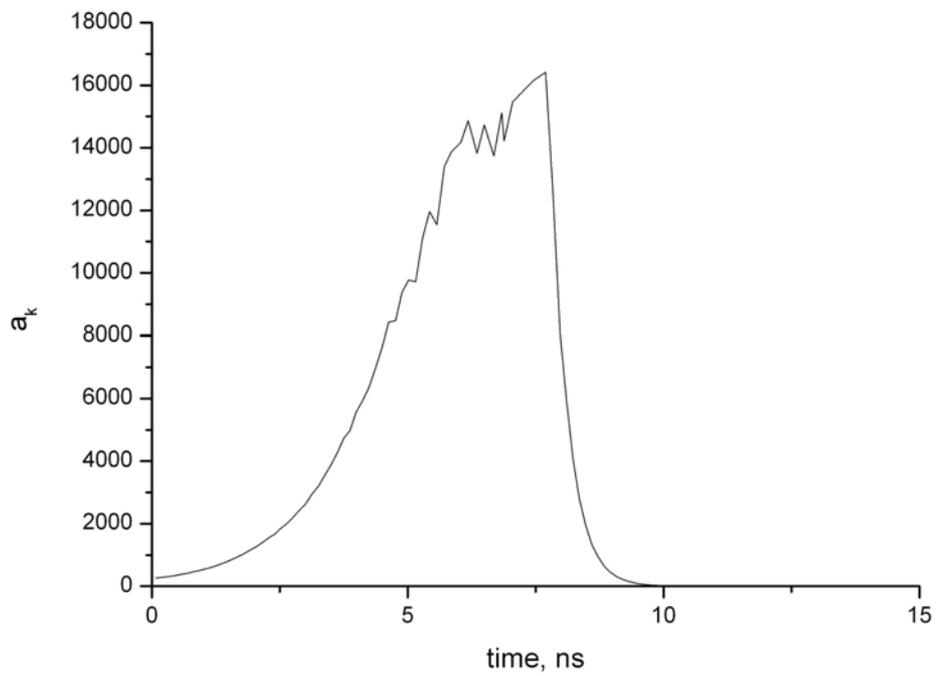

**Figure 5. Amplitude of the uniform mode as a function of time, no external field, current amplitude 0.76 mA.**

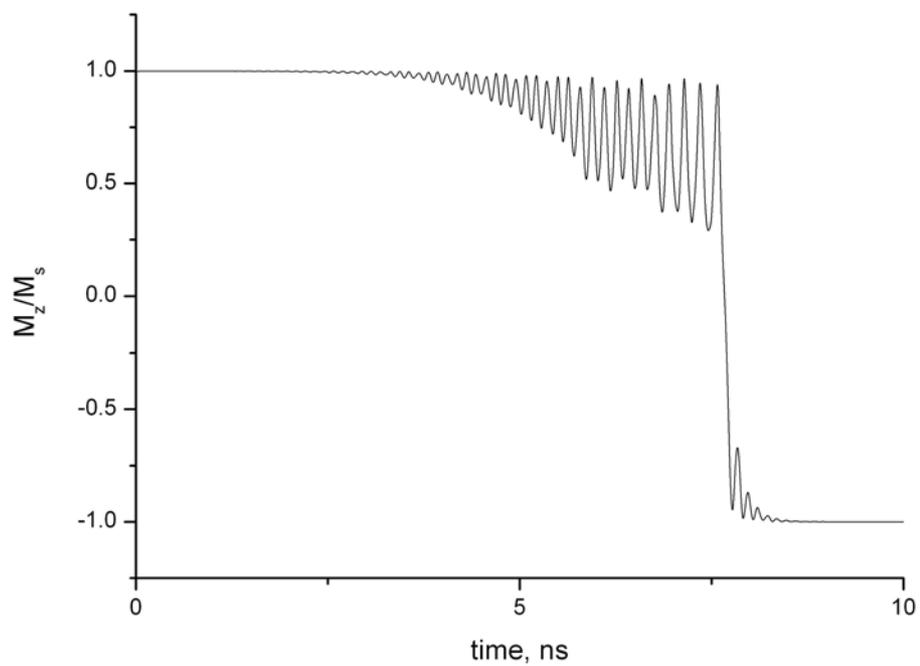

**Figure 6. Z component of the sample's magnetization as a function of time, no external field, current amplitude 0.76 mA.**

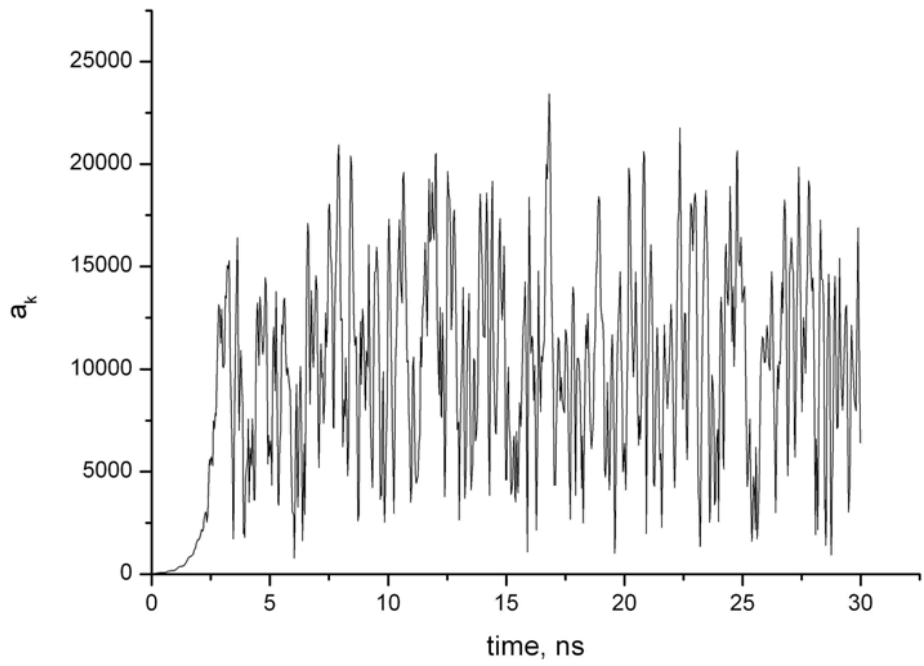

**Figure 7. Amplitude of the uniform mode as a function of time, $H_0$=2600 Oe, current amplitude 1.5 mA.**

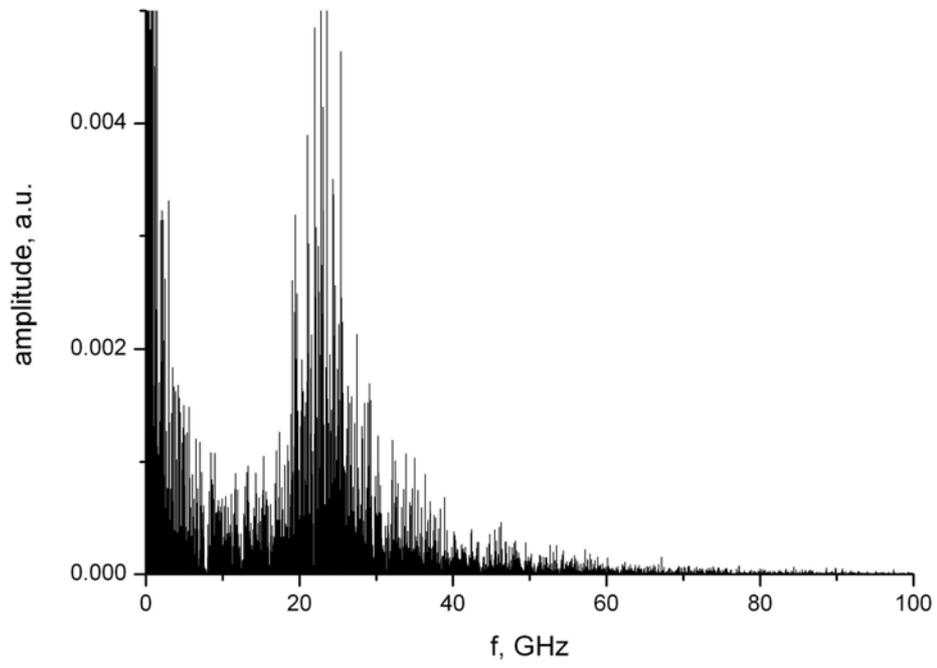

**Figure 8. Fourier transform of time dependency of $M_z$ component of magnetization of the free layer, 1.5 mA.**

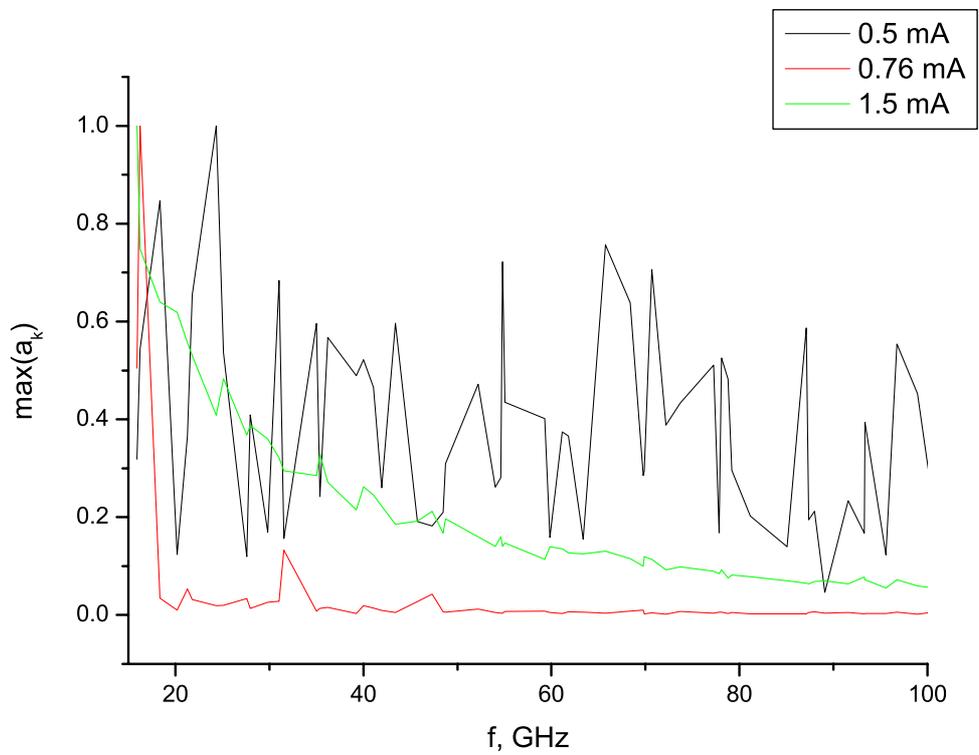

**Figure 9. Maximum amplitude of modes as a function of mode's frequency for different current values, $H_0$=2600 Oe.**

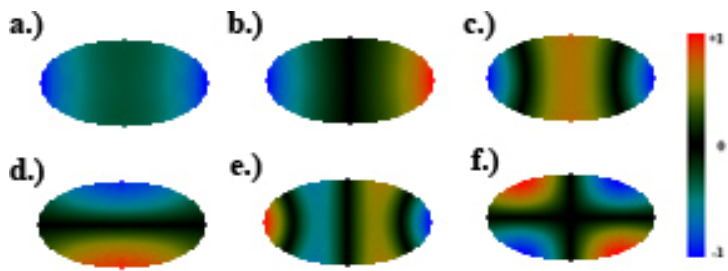

**Figure 10.** Z projection of the low frequency modes, $H_0$=2600 Oe: a.) f=15.91 GHz, b.) f = 16.25 GHZ, c.) f = 18.37 GHz, d.) f=20.18 GHz, e.) f=21.27 GHz, f.) f=21.79 GHz.